\documentclass[aps,pra,showpacs,groupedaddress,superscriptaddress,twocolumn]{revtex4}
\usepackage{amsfonts}
\usepackage{graphicx}
\usepackage{latexsym}
\usepackage{makeidx}
\usepackage{amsmath}
\usepackage{amssymb}
\usepackage{graphicx}
\usepackage{bm}

\def\bra#1{\mathinner{\langle{#1}|}}
\def\ket#1{\mathinner{|{#1}\rangle}}
\def\defeq{\mathrel{\mathop:}=}

\def\eqref#1{(\ref{#1})}

\begin{document}

\title{Universality of Sudden Death of Entanglement}
\author{F. F. Fanchini}
\email{fanchini@ifi.unicamp.br}
\affiliation{Departamento de F\'{\i}sica, Universidade Federal de Ouro Preto, CEP 35400-000, Ouro Preto, MG, Brazil}
\author{P. E. M. F. Mendon\c ca}
\affiliation{Divis\~{a}o de Ensino, Academia da For\c ca A\'{e}rea, CEP 13565-905, Caixa Postal 970, CEP 13643-970, Pirassununga, SP, Brazil}
\author{R. d J. Napolitano}
\affiliation{Instituto de F\'{\i}sica de S\~{a}o Carlos, Universidade de S\~{a}o Paulo, Caixa Postal 369, CEP 13560-970, S\~{a}o Carlos, SP, Brazil}

\begin{abstract}
We present a constructive argument to demonstrate the universality of the sudden death of entanglement in the case of two non-interacting qubits, each of which generically coupled to independent Markovian environments at zero temperature. Conditions for the occurrence of the abrupt disappearance of entanglement are determined and, most importantly, rigorously shown to be \emph{almost} always satisfied: Dynamical models for which the sudden death of entanglement does not occur are seen to form a highly idealized zero-measure subset within the set of all possible quantum dynamics.
\end{abstract}

\maketitle

\section{Introduction}
\noindent
The entanglement of quantum states is one of the most fundamental aspects of quantum mechanics and is crucial
for countless processes of quantum information and computation~\cite{93Bennett1895,97Bouwmeester575,04Riebe734,01Knill46,05Petta2180,01Raussendorf5188}. Recently, the study of entangled states perturbed by local environmental noise has become a subject of increasing interest. Unlike in decoherence processes~\cite{91Zurek36}, there are some entangled states which, depending on their couplings to the environment, become separable within a finite time interval~\cite{sudden0,sdmeas,07Almeida579,08Aolita080501,03Diosi157,04Dodd052105,08Salles022322,06Santos040305(R),07Yu459,06Yu140403,04Yu140404,rev}. Such states manifest what is called the sudden death of entanglement (SDE).

The incidence of SDE has been studied for a large class of initial entangled qubit states, and in the separate cases of amplitude-damping, depolarizing, dephasing, bit flipping, and bit-phase flipping~\cite{07Almeida579,08Aolita080501,08Salles022322,06Santos040305(R),07Yu459,06Yu140403,04Yu140404}. However, of all possible coupling agents inducing the noisy dynamics, the ones associated with these separate cases constitute very particular choices and, in many cases, not realistic ones.

As an example, consider the case where a photon, in a linear combination of zero- and one-photon states, traverses an optical medium and is scattered by an atom. In such a process, as it is well known, dispersion is positively correlated with absorption, implying that we cannot realistically describe the environmental noise as amplitude-damping alone; we must also consider dephasing.

Motivated by practical constraints such as above, and also by the theoretically large set of available coupling agents between the system and the environment, we are led to ask whether SDE is a general feature of quantum dynamics, or else a peculiarity of certain models. Although many authors have already argued in favor of the omnipresence of SDE \cite{zhang2009, cui2009, ikram2007, 08alqasimi, ali2009}, the present study conveys the first rigorous demonstration of the universality of this phenomenon in the case of two non-interacting qubits, each of which coupled to its own independent Markovian environment at zero temperature.

Our main result is obtained by first deriving conditions for the occurrence of SDE with generalized error agents, and then by proving that the set of coupling agents for which SDE may not occur has zero-measure with respect to the set of all possible couplings; meaning that, in the considered scenario, SDE is indeed a universal phenomenon.

The paper is organized as follows. In Section~\ref{sec:dynamics} we introduce the qubit interaction Hamiltonian for general error agents, write and solve the Born-Markov master equation for the dynamics of a single qubit, and review a procedure to construct the dynamics of two independent qubits from the single qubit dynamics. In Section~\ref{sec:SDEcond} we derive conditions for the occurrence of SDE, which are shown to be (almost) always met in Section~\ref{sec:universality}. We summarize our conclusions and some possible avenues for future research in Section~\ref{sec:conclusion}.

\section{Dynamics of one and two qubits under general error agents}\label{sec:dynamics}
\noindent
In this section, we derive an exact analytical expression for the dynamics of a single qubit coupled to a Markovian environment at zero temperature for a comprehensive class of interaction Hamiltonians. We also briefly review two known results that will be useful throughout. The first one describes how the dynamics of a system of two qubits --- each of which coupled to an independent environment --- can be constructed from the single-body dynamics, whereas the second one provides a recipe to quantify entanglement of a two-qubit system.

\subsection{The Lindblad equation}
\noindent
We proceed by calculating the dynamics of a single perturbed qubit, assuming a general interaction-picture Hamiltonian of the form
\begin{equation}
H_{I}(t)=({\bm \lambda}\cdot{\bm \sigma})\tilde{B}^\dagger(t)+({\bm \lambda}^\ast\cdot{\bm \sigma})\tilde{B}(t),\label{Hint}
\end{equation}
where ${\bm \lambda}$ is an arbitrary three-dimensional vector with complex components, $\tilde{B}(t)=U_E^\dagger(t)B U_E(t)$, ${\bm \sigma}={\bf \hat{x}}\sigma_{x}+{\bf \hat{y}}\sigma_{y}+{\bf \hat{z}}\sigma_{z}$, and $U_E(t)=\exp{(-iH_E t)}$, with $H_E$ representing the environment Hamiltonian (hereafter, we assume $\hbar=1$).

Without specifying $H_{E}$ and $\tilde{B}(t)$, the Born-Markov master equation for the Hamiltonian given above, at zero absolute temperature, $T=0$, can be derived by following the same steps outlined in Ref.~\cite{02Breuer}, and yields the Lindblad equation
\begin{eqnarray}
\!\!\!\!\!\!\!\frac{d{\rho}_{I}(t)}{dt}\! & = &\!\! -\gamma\left\{\frac{}{}\!\! \mathcal{L}^\dagger\mathcal{L}\rho_{I}(t)+\rho_{I}(t)\mathcal{L}^\dagger\mathcal{L}-2\mathcal{L}\rho_{I}(t)\mathcal{L}^\dagger\right\}\!,\label{lindblad}
\end{eqnarray}
with $\mathcal{L}\equiv{\bm \lambda}\cdot{\bm\sigma}$ and
\begin{eqnarray}
\gamma =\int _{-\infty}^{+\infty} ds \,\,{\rm Tr}_{E}\left\{\tilde{B}(0)\tilde{B}^{\dagger}(s)\rho _{E}(0)\right\}.\label{gamma}
\end{eqnarray}
To derive Eq.~\eqref{lindblad}, we suppose that $\rho _{E}(0)$ is stationary, that is, $[\rho _{E}(0),H_{E}]=0$, also that
\begin{eqnarray}
{\rm Tr}_{E}\left\{ \tilde{B}(t)\rho _{E}(0)\right\}=0\,,
\end{eqnarray}
and that $\tilde{B}(t)$ anihilates the ground state $\left|0\right\rangle $ of the environment:
\begin{eqnarray}
\tilde{B}(t)\left|0\right\rangle =0.
\end{eqnarray}
Moreover, we also assume that $\rho _{E}(0)=\left|0\right\rangle \!\left\langle 0\right|$ at $T=0$.

\subsection{Solution of the Lindblad equation}\label{sec:solution_lindblad}
\noindent
In order to solve Eq.~\eqref{lindblad}, it is convenient to
use the decomposition $\bm{\lambda}=\mathbf{u}+i\mathbf{v}$ with $\mathbf{u},\mathbf{v}\in{\mathbb{R}}^{3}$ and $|\mathbf{u}|^2+|\mathbf{v}|^2=1$ in the interaction Hamiltonian of Eq.~\eqref{Hint}, in such a way that the Lindbladian operator can be written as
\begin{equation}\label{eq:Ls}
\mathcal{L}=\left(\mathbf{u}+i\mathbf{v}\right)\cdot\bm{\sigma}\quad\mbox{and}\quad \mathcal{L}^\dagger\mathcal{L}=\mathbb{I}_2-2\mathbf{w}\cdot\bm{\sigma}\,,
\end{equation}
where $\mathbf{w}\defeq\mathbf{u}\times\mathbf{v}$.

Substituting the above into Eq.~\eqref{lindblad}, and expressing $\rho_I(t)$ in terms of its Bloch vector $\mathbf{r}(t)$,
\begin{equation}
\rho_I(t)=\frac{\mathbb{I}_2}{2}+\frac{1}{2}{\mathbf r}(t)\cdot\bm{\sigma}\,,\label{rhobloch}
\end{equation}
we obtain the following master equation for $\mathbf{r}(t)$, equivalent to Eq.~\eqref{lindblad} for $\rho_I(t)$,
\begin{equation}
\frac{d{\mathbf r}(t)}{dt} = 4\gamma\left\{\frac{}{}\mathbf{u}\left[{\mathbf u}\cdot{\mathbf r}(t)\right]+\mathbf{v}\left[\mathbf{v}\cdot\mathbf{r}(t)\right]+2\,\mathbf{w}-\mathbf{r}(t)\right\}\,.\label{drdt}
\end{equation}

In what follows, the solution of Eq.~\eqref{drdt} is given in two separate cases, depending on whether $\mathbf{u}$ and $\mathbf{v}$ are chosen to be linearly-dependent or independent vectors.

\subsubsection{Linear Dependence: Flip Hamiltonians}
\noindent
If $\mathbf{u}$ and $\mathbf{v}$ are chosen to be \emph{linearly-dependent}, then we have $\mathbf{w}=0$, in which case it is not difficult to show that the solution of Eq.~\eqref{drdt} is
\begin{equation}\label{eq:rLDu}
\mathbf{r}(t)=\mathrm{e}^{-4\gamma t}\mathbf{r}_0+\left(1-\mathrm{e}^{-4\gamma t}\right)\left(\mathbf{r}_0\cdot{\bf \hat{u}}\right){\bf \hat{u}}\,,
\end{equation}
where we have defined ${\mathbf r}_0\equiv {\mathbf r}(0)$, and the unity vector ${\bf \hat{u}}$ such that $\mathbf{u}=|\mathbf{u}|{\bf \hat{u}}$.

Alternatively, it is relatively straightforward to represent the solution above in the operator-sum form,
\begin{equation}
\rho_I(t)=\sum_i K_i(t) \rho_I(0) K_i^\dagger(t)\,,
\end{equation}
with Kraus operators given by
\begin{equation}\label{eq:KOflip}
K_1(t)=\sqrt{p(t)}\;\mathbb{I}_2\quad\mbox{and}\quad K_2(t)=\sqrt{1-p(t)}\;(\mathbf{\hat{u}}\cdot\bm{\sigma})
\end{equation}
where $p(t)=(1+\mathrm{e}^{-4\gamma t})/2$.

Noticeably, the Kraus operators above are identical to those of the bit flip, bit-phase flip and phase flip channels if
$\mathbf{\hat{u}}$ is chosen equal to $\mathbf{\hat{x}}$, $\mathbf{\hat{y}}$ or $\mathbf{\hat{z}}$, respectively~\cite{00Nielsen}, which motivates the name \emph{flip Hamiltonian}. In fact, any interaction Hamiltonian arising from a linear dependent choice of $\mathbf{u}$ and $\mathbf{v}$ yields a type of ``flip channel''. The effect of an arbitrary flip channel on a single qubit is detailed in the Appendix~\ref{app:inbuilt}. In that appendix, we also illustrate the action of more general dissipative Hamiltonians on the Bloch ball.

\subsubsection{Linear Independence: Dissipative Hamiltonians}\label{sec:dissipham}
\noindent
In this case, since $\mathbf{w}\neq 0$, the Bloch vector can be expressed as
\begin{eqnarray}
\mathbf{r}(t) & = & f(t)\mathbf{u}+g(t)\mathbf{v}+h(t)\mathbf{w}\,, \label{rbloch}
\end{eqnarray}
and Eq.~\eqref{drdt} is solved with the following choices of $f(t)$, $g(t)$, and $h(t)$:
\begin{eqnarray}
h(t) & = &2 - \left(2-\frac{{\mathbf r}_0\cdot\mathbf{w}}{\left|\mathbf{w}\right|^{2}}\right)\mathrm{e}^{-4\gamma t}\,, \label{h}
\end{eqnarray}
and
\begin{eqnarray}\hspace{-0.6cm}\left[\begin{array}{c}
f\left(t\right)\\
g\left(t\right)\end{array}\right]&\!\!\! =\!\! & \mathcal{M}\left[\begin{array}{c}
\left(\mathbf{v}\times\mathbf{w}\right)\cdot\mathbf{r}_0\\
\left(\mathbf{w}\times\mathbf{u}\right)\cdot\mathbf{r}_0\end{array}\right]\frac{\mathrm{e}^{-2\gamma t}}{{\left|\mathbf{w}\right|^2}},\label{fg}
\end{eqnarray}
with the $2 \times 2$ matrix $\mathcal{M}$ given by
\begin{equation}
\mathcal{M}=\cosh(2\gamma q\, t)\,\mathbb{I}_2\,+\sinh(2\gamma q\,t)\left[\frac{(\mathbf{u}\cdot\mathbf{v})\sigma_{x}+\chi\sigma_{z}}{q}\right],\label{M}
\end{equation}
where $q = { {\sqrt{{\chi}^{2}+\left(\mathbf{u}\cdot\mathbf{v}\right)^2}}}$ and $\chi  = \left(|\mathbf{u}|^2-|\mathbf{v}|^2\right)/2$.

From the equations above, a set of Kraus operators $\{K_i\}$ for the dissipative dynamics of a single qubit can be obtained following a standard procedure \cite{02Verstraete,05Salgado55} outlined here in the Appendix~\ref{app:Kraus}. In summary, Eqs.~\eqref{eq:rLDu}-\eqref{M} describe the dynamics of a qubit coupled to a Markovian environment at zero temperature for a broad class of interaction Hamiltonians parametrized by $\bm{\lambda}$.

\subsection{Asymptotic limit}\label{sec:asymptotics}
\noindent
In order to enable the study of the entanglement dynamics of a two-qubit system in Section~\ref{sec:SDEcond}, it will be helpful to determine the asymptotic form of the Bloch vector of each individual qubit state, evolving under flip and dissipative Hamiltonians.

In the case of flip Hamiltonians, each individual Bloch vector converges to $\mathbf{r}_\infty=\left({\mathbf r}_0\cdot\mathbf{\hat{u}}\right)\mathbf{\hat{u}}$ [cf.~Eq.~\eqref{eq:rLDu}], revealing that the final state depends both on the interaction Hamiltonian (through $\mathbf{\hat{u}}$) and on the initial condition (through $\mathbf{r}_0$).

In contrast, for dissipative Hamiltonians, the final state depends only on the interaction Hamiltonian. In fact, a quick inspection of Eqs.~\eqref{rbloch}-\eqref{fg} shows that each individual Bloch vector converges to $\mathbf{r}_\infty=2\mathbf{w}$. The independence of $\mathbf{r}_\infty$ on the initial state conveys important information about the SDE phenomenon in the case of dissipative dynamics.

\subsection{From a single qubit to two independent qubits}\label{sec:bellomo}
\noindent
Since we are considering that the two qubits are independent and interact only locally with their reservoirs, their dynamics can be obtained by the set of Kraus operators that act over each single body \cite{yu03}.
Accordingly, in the case of a two-qubit initial state $\rho(0)$, we have the following Kraus decomposition of $\rho(t)$:
\begin{equation}\label{eq:bellomo}
\rho(t)=\left[K_{i}^{(1)}(t)\otimes K_{j}^{(2)}(t)\right]\rho(0)\left[{K_{i}^{(1)}}(t)\otimes {K_{j}^{(2)}}(t)\right]^\dagger
\end{equation}
where repeated indexes are to be summed over, and the corresponding matrices $K_{i}^{(n)}$ are Kraus operators describing each single qubit dynamics.

\subsection{Quantifying Entanglement}\label{sec:quant_ent}
\noindent
To examine the entanglement dynamics of the density matrix $\rho(t)$ defined above, we will utilize the concurrence~\cite{97Hill5022,98Wootters2245}, hereafter denoted by $C[\rho(t)]$. According to this measure, $\rho(t)$ is separable if and only if $C[\rho(t)]=0$. On the other hand, the more entangled $\rho(t)$ is, the larger is $C[\rho(t)]$.

The concurrence is defined as
 \begin{equation}
C[\rho(t)]=\max{[0,\Lambda(t)]}\,,
 \end{equation}
where $\Lambda(t)=l_1-l_2-l_3-l_4$, and $l_1\geq l_2\geq l_3\geq l_4$ are the square roots of the eigenvalues of the matrix
$\rho(t)\sigma_{y}^{\otimes 2}\rho^*(t)\sigma_{y}^{\otimes 2}$, where the complex conjugate of $\rho(t)$, denoted $\rho^*(t)$, is taken with respect
to the computational basis.

Given an initially entangled state, we note that a \emph{sufficient condition} for the occurrence of SDE is that $\Lambda(t\to\infty)$ --- henceforth denoted $\Lambda_\infty$ --- is a strictly negative number. In this case, we can guarantee that $\Lambda(\tau)=0$ for some finite time $\tau$, or in other words, that all the initial entanglement disappeared at $\tau$. Moreover, since the concurrence is known to decay monotonically under the Markovian approximation~\cite{08Dajka042316}, in our analysis there will be no room for entanglement revival after $\tau$.

In the next section, the above sufficiency criterion will be central in deriving more immediate conditions for the occurrence of SDE. Our goal is to obtain criteria that do not require the knowledge of the asymptotic state, but instead that predict the incidence of SDE from the simple observation of the initial states and/or the interaction Hamiltonian.

\section{Conditions for SDE}\label{sec:SDEcond}
\noindent
In this section, we prove two theorems that characterize the occurrence of SDE in terms of
quantities which are generally assumed to be known \emph{a priori}: the initial state of the system and the Hamiltonian mediating the interaction between the system and the environment.

Our first theorem deals with the case of flip Hamiltonians, and can be seen as a generalization of a previous result by Huang and Zhu~\cite{07Huang062322}. The second theorem, which treats the case of more general dissipative Hamiltonians, had not yet been identified to the best of our knowledge. In the Appendix~\ref{app:watch}, this second theorem is illustrated through the explicit analysis of some particular cases of interest.

\vspace*{12pt}
\noindent
{\bf Theorem~1.}
{\it A necessary and sufficient condition for the occurrence of SDE in a system of two entangled qubits evolving independently under flip Hamiltonians, is that the matrix
\begin{equation}
\widetilde{\rho}_0=(U^{(1)}\otimes U^{(2)})^\dagger \rho_0 (U^{(1)}\otimes U^{(2)})
\end{equation}
has no zeroes in the diagonal. In the above, $\rho_0$ is the initial two-qubit density matrix, and $U^{(n)}=\mathbb{I}_2$ if $\mathbf{\hat{u}}^{(n)}=\mathbf{\hat{z}}$, or else
\begin{equation}\label{eq:unitary}
U^{(n)}=\left[\begin{array}{cc}
\cos\varphi^{(n)} & -\frac{\left[u_1^{(n)}-i u_2^{(n)}\right]}{\sqrt{1-\left[u_3^{(n)}\right]^2}}\sin\varphi^{(n)}\\
\frac{\left[u_1^{(n)}+i u_2^{(n)}\right]}{\sqrt{1-\left[u_3^{(n)}\right]^2}}\sin\varphi^{(n)}&\cos\varphi^{(n)}
\end{array}\right]\,,
\end{equation}
 where $\varphi^{(n)}=\frac{1}{2}\arccos{u_3^{(n)}}$ and $[u_1^{(n)}, u_2^{(n)}, u_3^{(n)}]=\mathbf{\hat{u}}^{(n)}$}.

\vspace*{12pt}
\noindent
{\bf Proof.} We start by showing that, for purposes of entanglement quantification, an arbitrary local flip dynamics of $\rho_0$ is equivalent to local dephasing of $\widetilde{\rho}_0$.

First, it is easy to check that $U^{(n)}$ is a unitary matrix such that
\begin{equation}\label{eq:unitary_vector}
U^{(n)} \sigma_z {U^{(n)}}^\dagger =  \mathbf{\hat{u}}^{(n)}\cdot\bm{\sigma}\,.
\end{equation}
Hence, the Kraus operators of an arbitrary flip channel $K_1^{(n)}$ and $K_2^{(n)}$ [cf. Eq.~\eqref{eq:KOflip}] can be related to the Kraus operators of a phase-flip channel $\widetilde{K}_1^{(n)}$ and $\widetilde{K}_2^{(n)}$ as follows:
\begin{equation}
K_1^{(n)}=U^{(n)} \widetilde{K}_1^{(n)} {U^{(n)}}^\dagger\quad\mbox{and}\quad K_2^{(n)}=U^{(n)} \widetilde{K}_2^{(n)} {U^{(n)}}^\dagger\,.
\end{equation}
The two-qubit density matrix dynamics is obtained by plugging the initial state $\rho_0$ and the Kraus operators above into Eq.~\eqref{eq:bellomo}, which becomes
\begin{equation}
\rho(t)=\left[U^{(1)}\otimes U^{(2)}\right]\widetilde{\rho}(t)\left[U^{(1)}\otimes U^{(2)}\right]^\dagger
\end{equation}
with
\begin{equation}
\widetilde{\rho}(t)=\left[\widetilde{K}_i^{(1)}\otimes\widetilde{K}_j^{(2)}\right]\widetilde{\rho}_0 \left[\widetilde{K}_i^{(1)}\otimes\widetilde{K}_j^{(2)}\right]^\dagger
\end{equation}
where the Einstein convention sum is adopted.

From the equations above, it is clear that the task of quantifying the entanglement of $\rho(t)$ is equivalent to the task of quantifying the entanglement of $\widetilde{\rho}(t)$, since $\rho(t)$ and $\widetilde{\rho}(t)$ are related by a simple \emph{local} unitary transformation, and thus have the same amount of entanglement. In other words, the problem of quantifying the entanglement of $\rho_0$ evolving under arbitrary flip channels is equivalent to the problem of quantifying the entanglement of the ``rotated density matrix'' $\widetilde{\rho}_0$, evolving under dephasing channels.

With this equivalence in mind, the theorem is promptly proved by invoking an independent result due to Huang and Zhu~\cite{07Huang062322}, which states that for a two-qubit system whose parties evolve independently under the influence of dephasing channels, the absence of zeros in the diagonal of the initial density matrix is a necessary and sufficient condition for the occurrence of SDE. In what follows, we provide an independent proof of the sufficiency clause by showing that $\Lambda_\infty\leq 0$ (cf. Section~\ref{sec:quant_ent}), with saturation only if $\widetilde{\rho}_0$ has at least one zero in the diagonal.

Let $r_{i,j}$, with $1 \leq i,j\leq 4$, be the matrix elements of $\widetilde{\rho}_0$. A direct computation shows that $\widetilde{\rho}(t\to\infty)$ is simply the matrix obtained from $\widetilde{\rho}_0$ by zeroing the off-diagonal elements and keeping the diagonal ones. From the resulting diagonal matrix, it is easy to find that $\Lambda_\infty=-2\min(r_{1,1} r_{4,4}\,,\,r_{2,2} r_{3,3})\leq 0$, where the inequality follows from the fact that a positive semidefinite matrix (such as $\widetilde{\rho}_0$) can only have non-negative diagonal entries. Moreover, it is clear that saturation can only occur if $r_{1,1} r_{4,4}=0$ or $r_{2,2} r_{3,3}=0$.\hspace{0.3cm}.

In what follows we provide analogous conditions for the occurrence of SDE in the case of dissipative Hamiltonians. Before stating and proving our conditions, it will be helpful to present the following lemma

\vspace*{12pt}
\noindent
{\bf Lemma~1.}
{\it If a set of Kraus operators $\{K_i^{(n)}\}$ induces a quantum channel such that
\begin{equation}\label{eq:Kon1}
\sum_i K_i^{(n)}\varrho_2 {K_i^{(n)}}^\dagger=\frac{\mathbb{I}_2}{2}+\mathbf{w}^{(n)}\cdot\bm{\sigma}
\end{equation}
for an arbitrary qubit density matrix $\varrho_2$, and where $\mathbf{w}^{(n)}$ is implicitly and exclusively defined by the Kraus operators [i.e., $\varrho_2$ is independent of $\mathbf{w}^{(n)}$], then
\begin{eqnarray}\label{eq:Kon2}
\sum_{i,j}\left(K_i^{(1)}\otimes K_j^{(2)}\right)\varrho_4 \left(K_i^{(1)}\otimes K_j^{(2)}\right)^\dagger\nonumber\\=\left[\frac{\mathbb{I}_2}{2}+\mathbf{w}^{(1)}\cdot\bm{\sigma}\right]\otimes\left[\frac{\mathbb{I}_2}{2}+\mathbf{w}^{(2)}\cdot\bm{\sigma}\right]
 \end{eqnarray}
 for every $4\times 4$ density matrix $\varrho_4$.}

\vspace*{12pt}
\noindent
{\bf Proof.}
First, notice that the linearity of Eq.~\eqref{eq:Kon1} and a few appropriate choices of $\varrho_2$, enables one to determine the action of the quantum channel on the operator basis $\{\ket{\mu}\!\bra{\nu}\}_{\mu,\nu\in\{0,1\}}$, namely
\begin{equation}\label{eq:Konbasis}
\sum_i K_i^{(n)} \ket{\mu}\!\bra{\nu} {K_i^{(n)}}^\dagger=\delta_{\mu,\nu}\left(\frac{\mathbb{I}_2}{2}+\mathbf{w}^{(n)}\cdot\bm{\sigma}\right)\,.
\end{equation}
Now, expand $\varrho_4$ in the corresponding tensor product basis
\begin{equation}
\varrho_4=\sum_{\alpha,\beta,\kappa,\gamma} p_{\alpha,\beta,\kappa,\gamma} \ket{\alpha}\!\bra{\beta}\otimes\ket{\kappa}\!\bra{\gamma}
\end{equation}
where the expansion coefficients must satisfy $\sum_{\alpha,\kappa}p_{\alpha,\alpha,\kappa,\kappa} = 1$ in order to guarantee the unity-trace property of $\varrho_4$. The lemma is trivially proved by plugging the expansion above into the left-hand side of Eq.~\eqref{eq:Kon2}, and by evaluating the resulting expression with the aid of Eq.~\eqref{eq:Konbasis}.

\vspace*{12pt}
\noindent
{\bf Theorem~2.}
{\it A sufficient condition for the occurrence of SDE in a system of two entangled qubits evolving independently under dissipative Hamiltonians, is that the pair of vectors $\mathbf{u}^{(n)}$ and $\mathbf{v}^{(n)}$ specifying each Hamiltonian satisfies
\begin{equation}\label{eq:dissip_SDE}
|\mathbf{u}^{(n)}\times\mathbf{v}^{(n)}|\neq\frac{1}{2}\,,
\end{equation}
for $n=1,2$. Or, equivalently, none of the qubits undergoes a standard amplitude-damping channel.}

\vspace*{12pt}
\noindent
{\bf Proof.}
Recall from Section~\ref{sec:asymptotics} that, in the asymptotic regime, a single qubit evolving under a dissipative Hamiltonian reaches a state which depends only on the interaction Hamiltonian via the vector $\mathbf{w}$. For two qubits, each of which evolving under independent dissipative Hamiltonians, the asymptotic global state can be promptly obtained by invoking Eq.~\eqref{eq:bellomo} and the result of Lemma~1:
\begin{equation}\label{eq:rhoinfdiag}
\rho_\infty=\left[\frac{\mathbb{I}_2}{2}+\mathbf{w}^{(1)}\cdot\bm{\sigma}\right]\otimes\left[\frac{\mathbb{I}_2}{2}+\mathbf{w}^{(2)}\cdot\bm{\sigma}\right]\,,
\end{equation}
which implies that any initial entanglement of $\rho_0$ has disappeared. In what follows, we focus on determining some cases where such entanglement disappearance occurred abruptly (SDE), as opposed to asymptotically.

Substituting the trivial identities $\mathbf{w}^{(n)}=|\mathbf{w}^{(n)}|\mathbf{\hat{w}}^{(n)}$ and $\mathbf{\hat{w}}^{(n)}\cdot\bm{\sigma} = W^{(n)} \sigma_z {W^{(n)}}^\dagger$ into Eq.~\eqref{eq:rhoinfdiag} [in which the unitary matrix $W^{(n)}$ is defined in analogy to Eqs.~\eqref{eq:unitary} and~\ref{eq:unitary_vector}], one finds that
\begin{equation}
\rho_\infty=\left[W^{(1)}\otimes W^{(2)}\right]\widetilde{\rho}_\infty\left[W^{(1)}\otimes W^{(2)}\right]^\dagger
\end{equation}
with
\begin{equation}
\widetilde{\rho}_\infty=\left[\frac{\mathbb{I}_2}{2}+|\mathbf{w}^{(1)}|\sigma_z\right]\otimes\left[\frac{\mathbb{I}_2}{2}+|\mathbf{w}^{(2)}|\sigma_z\right]\,.
\end{equation}
Thus, without loss of generality, the task of quantifying the entanglement of $\rho_\infty$ can be replaced with the task of quantifying the entanglement of the \emph{diagonal} density matrix $\widetilde{\rho}_\infty$, for which one can easily compute
\begin{equation}
\Lambda_\infty=-\frac{1}{2}\sqrt{\left[1-|2\mathbf{w}^{(1)}|^2\right]\left[1-|2\mathbf{w}^{(2)}|^2\right]}\leq 0\,.
\end{equation}
In the above, the inequality is trivially satisfied with saturation if $|\mathbf{w}^{(n)}|=1/2$ for $n=1$ or $n=2$. As noted before (cf. Section~\ref{sec:quant_ent}), a strictly negative
value of $\Lambda_\infty$ indicates the disappearance of entanglement in a finite time interval, leading to the sufficient condition for SDE given by Eq.~\eqref{eq:dissip_SDE}. Remarkably, this is not a necessary condition since $\Lambda_\infty$ may have vanished at a finite time and remained zero ever since (cf. Appendix~\ref{app:watch}).

Finally, we note that if $|\mathbf{w}^{(n)}|=1/2$, then the $n$-th qubit collapses to a pure state, regardless of the initial preparation. As seen in Appendix~\ref{app:inbuilt}, this is precisely what is accomplished by a standard amplitude damping channel. Therefore, with this only exception, SDE occurs for all possible choices of dissipative Hamiltonians.

\section{Universality of SDE}\label{sec:universality}
\noindent
With the aid of the conditions for SDE derived in the preceding section, we are now
ready to prove our main result.

\vspace*{12pt}
\noindent
{\bf Theorem~3.}
{\it In the case of two non-interacting qubits, each of which coupled to independent
Markovian environments at zero temperature, the set of dynamics for which SDE is not
guaranteed to occur has zero-measure in the set of all possible dynamics.
In other words, SDE is universal for such systems.}

\vspace*{12pt}
\noindent
{\bf Proof.}
For a single qubit, the set of all possible dynamics generated by the Hamiltonian of Eq.~\eqref{Hint} is equivalent to the set
\begin{equation}
\Omega=\{(\mathbf{u}, \mathbf{v})\;:\; \mathbf{u}, \mathbf{v}\in \mathbb{R}^3, |\mathbf{u}|^2+|\mathbf{v}|^2=1\}\,,
\end{equation}
thanks to the parametrization $\bm{\lambda}=\mathbf{u}+i\mathbf{v}$ with $|\bm{\lambda}|=~1$.

The dimension of a set is defined as the number of independent parameters needed to
describe a point in the set, in which case $\dim\Omega=5$, since any element of $\Omega$ can be written in terms of the parameters $R\in[0,1]$, $\theta,\theta^\prime\in[0,2\pi]$ and $\phi,\phi^\prime\in[0,\pi]$ as follows:
\begin{eqnarray}\label{eq:paramuv}
\mathbf{u}&=&R\left(\cos\theta\cos\phi\;\mathbf{\hat{x}}+\sin\theta\cos\phi\;\mathbf{\hat{y}}+\sin\phi\;\mathbf{\hat{z}}\right)\\
\mathbf{v}&=&\sqrt{1-R^2}\left(\cos\theta^\prime\cos\phi^\prime\;\mathbf{\hat{x}}+\sin\theta^\prime\cos\phi^\prime\;\mathbf{\hat{y}}+\sin\phi^\prime\;\mathbf{\hat{z}}\right)\nonumber
\end{eqnarray}

We now show that the identified dynamical models for which SDE does not necessarily occur form sets of dimension $3$. Let
\begin{eqnarray}
\Omega_{F}&=&\Omega \cap \{(\mathbf{u}, \mathbf{v})\; :\; \mathbf{v}=\alpha\mathbf{u}, \alpha\in\mathbb{R}\}\,,\\
\Omega_{AD}&=&\Omega \cap \{(\mathbf{u}, \mathbf{v})\;:\; |\mathbf{u}\times\mathbf{v}|=\frac{1}{2}\}\,,
\end{eqnarray}
where $\Omega_{F}$ is equivalent to the set of flip Hamiltonians, and $\Omega_{AD}$ is equivalent to the set of Hamiltonians producing standard amplitude damping channels (cf. Appendix~\ref{app:inbuilt}).

A straightforward computation shows that any element of $\Omega_F$ can be parametrized as in Eq.~\eqref{eq:paramuv} with $R=(1+\alpha^2)^{-1/2}$, $\theta^\prime=\theta$ and $\phi^\prime=\phi$, in such a way that two free parameters are eliminated. As a result, we have $\dim\Omega_F=3$. Similarly, one can easily  show that the elements of $\Omega_{AD}$ can be parametrized as in Eq.~\eqref{eq:paramuv} with $R=1/\sqrt{2}$, and one of the angles fixed by the solution of the transcendental equation $\mathbf{u}\cdot\mathbf{v}=0$. As a result, we also have $\dim\Omega_{AD}=3$.

Since the volume of $\Omega_F$ and $\Omega_{AD}$ with respect to the standard Lebesgue measure in $\mathbb{R}^5$ vanishes, the theorem is proved. Pictorially, one should consider the analogous lower dimensional scenario of a curve lying inside a solid block: it is impossible to cover the whole block with a countable number of curves, so the curve has zero-measure with respect to the block. The same argument holds for higher dimensions.

\section{Concluding Remarks}\label{sec:conclusion}
\noindent
As far as nature does not present any preferential way non-interacting qubits can couple to their assumed-Markovian environments, SDE is almost-surely universal even at zero temperature. The universality achieved in the present work is conditional to the stated assumptions which, by their experimental banality, are to be commonsensically deemed ubiquitous, except for the zero-temperature restriction. Nonetheless, recent theoretical investigations demonstrate that, at finite temperatures, a relevant set of two-qubit entangled states, robust against SDE at zero temperature, get disentangled at a finite time \cite{08alqasimi}. The results of those investigations show that if SDE occurs at low temperatures, it necessarily occurs at higher temperatures too.

In what concerns extensions of the present paper, it would be
interesting to verify whether or not SDE is also universal in the 
more general framework of two interacting qudits. In this case, the key 
difficulties to be confronted by an approach such as ours are two-fold:

First, entanglement measures for arbitrary dimensional systems are
still an active field of research, so far without the recognition of a 
measure which is both easy to compute (in an analytical sense) and well
motivated from a physical perspective \cite{reports}. In addition,
even if such a measure were identified, a study of SDE based on it would
probably be too advanced for comparison with experimental
realization, given the current difficulties arising in the
experimental quantification of entanglement \cite{qip}. Because of these,
a first attempt to extend the results of the present work should focus on
a system consisting of a qubit and a qutrit, in which case entanglement 
could be easily calculated and experimentally quantified by means of the negativity \cite{peres, horo}.

The second drawback relates to the Hamiltonian modeling of the
coupling between larger dimensional systems and the environment,
something which, to the best of our knowledge, has not been
accomplished so far (not even for qutrit states).  Furthermore, the
case of non-Markovian environments at finite temperature could also be
pointed out as another generalization of the present work.

\section*{Acknowledgments}
This work has been supported by Funda\c{c}\~{a}o de Amparo \`{a} Pesquisa do Estado de S\~{a}o Paulo, Brazil, project number 05/04105-5, the Brazilian National Institute of Science and Technology for Quantum Information (INCT-IQ), and the Millennium Institute for Quantum Information -- Conselho Nacional de Desenvolvimento Cient\'{\i}fico e Tecnol\'{o}gico, Brazil.+
\newpage


\appendix
\section{In-built standard dynamics}\label{app:inbuilt}
In this appendix, we illustrate the generality of our noise model by showing that it encompasses various
standard single qubit dynamics. This appendix is divided in two parts: First, we show that flip Hamiltonians lead to ``flip channels'' such as bit flip, phase flip and bit-phase flip. In the second part, we show that dissipative Hamiltonians produce dynamics represented by channels such as depolarizing and amplitude-damping.

\subsection*{Flip Channels}
\begin{figure} [htbp]
\includegraphics[width=8.5cm]{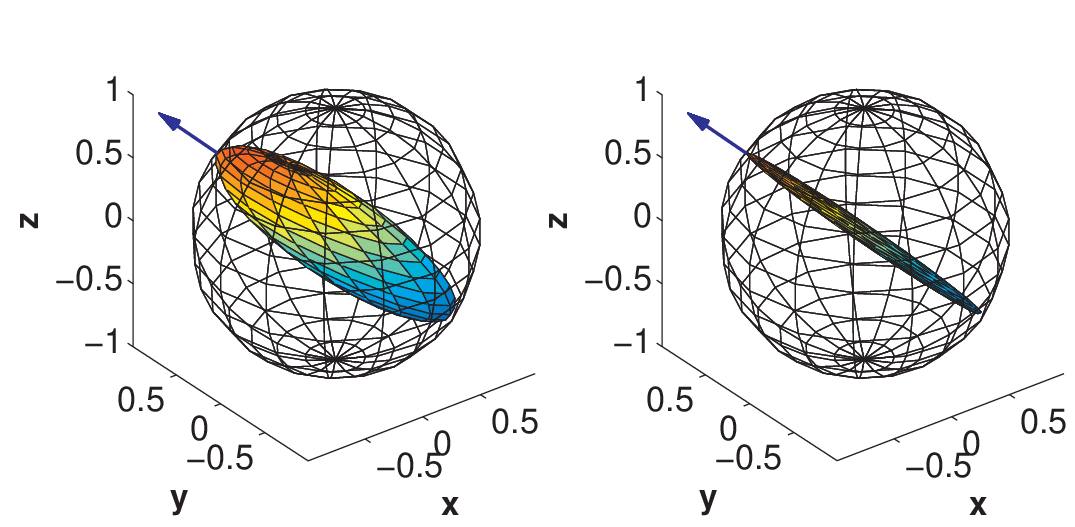}
\caption{\label{fig:flip_ch} In the first Bloch sphere, $\gamma t = 0.3$, while in the second, $\gamma t = 0.7$. The arrow designates the direction $\mathbf{\hat{u}}$.}
\end{figure}

\noindent
In Eq.~\eqref{eq:rLDu}, we found that the Bloch vector of a single qubit evolving under a flip Hamiltonian always points towards
a direction ${\bf \hat{u}}\in \mathbb{R}^3$ specified by the Hamiltonian. Now let
${\bf \hat{u}}^\perp$ and ${\bf \hat{u}}^\vdash$ be orthogonal directions to ${\bf \hat{u}}$ such that $\{{\bf \hat{u}},{\bf \hat{u}}^\perp,{\bf \hat{u}}^\vdash\}$ forms an orthonormal basis for $\mathbb{R}^3$. We find that the component-wise evolution of $\mathbf{r}(t)$ satisfies
\begin{eqnarray}\label{eq:dyn_comp_LD}
\mathbf{r}(t)\cdot{\bf \hat{u}}&=&\mathbf{r}_0\cdot{\bf \hat{u}}\label{eq:dyn_comp_LDa}\\
\mathbf{r}(t)\cdot{\bf \hat{u}}^\perp&=&\mathrm{e}^{-4\gamma t}\left(\mathbf{r}_0\cdot{\bf \hat{u}}^\perp\right)\\
\mathbf{r}(t)\cdot{\bf \hat{u}}^\vdash&=&\mathrm{e}^{-4\gamma t}\left(\mathbf{r}_0\cdot{\bf \hat{u}}^\vdash\right)
\end{eqnarray}
which shows that while the ${\bf \hat{u}}$-component remains unchanged, the orthogonal components exponentially decay in time at a rate $4\gamma$. The effect of such a channel on the whole Bloch sphere can be visualized in Fig.~\ref{fig:flip_ch} for two time-steps and an arbitrarily chosen direction ${\bf \hat{u}}$, designated by the arrow.

From the figure, it is clear that, asymptotically, the Bloch ball collapses into a diameter of the original sphere. As it should be expected, this observation is consistent with the result of Sec.~\ref{sec:asymptotics}, which establishes that $\mathbf{r}_\infty$ is not completely specified by this type of interaction Hamiltonian, but also by $\mathbf{r}_0$. That is because different initial states can collapse to different points of a particular diameter (specified by ${\bf \hat{u}}$). Furthermore, it is obvious from Fig.~\ref{fig:flip_ch} that if ${\bf \hat{u}}$ were chosen to be one of the Cartesian versors ${\bf \hat{x}}$, ${\bf \hat{y}}$, or ${\bf \hat{z}}$, then the resulting channel would be, respectively, a bit flip, a bit-phase flip, or a phase flip.

Notice that Fig.~\ref{fig:flip_ch} illustrates the most general dynamics that can arise from a flip Hamiltonian. Nevertheless, this is certainly not the most general dynamics that there is. For example, in the present case the Bloch ellipsoid will always be centered at the origin (\emph{unital} channel), and will always touch the Bloch sphere in two antipodal points, meaning that there is always a pair of orthogonal pure states that remain pure even after coupling to the environment. Needless to say, these features are not generally observed in quantum dynamics.

Next, we show that dissipative Hamiltonians give rise to a whole new class of channels, including non-unital channels such as the amplitude-damping, and unital channels that globally drop purity, such as the depolarizing channel.

\subsection*{Amplitude-Damping and Depolarizing Channels}
\noindent
Here, we take $\bm{\lambda}=\mathbf{u}+i\mathbf{v}$ with $\mathbf{u}$ and $\mathbf{v}$ \emph{linearly-independent} vectors. In particular,
we choose
\begin{eqnarray}
\mathbf{u}&=&{\bf \hat{x}}\sin\phi\cos\theta+{\bf \hat{z}}\cos\phi\label{eq:egu}\\
\mathbf{v}&=&-{\bf \hat{y}}\sin\phi\sin\theta\label{eq:egv}
\end{eqnarray}
in which case the normalization condition $|\bm{\lambda}|=1$ is fulfilled for any $\theta \in [0,2\pi]$ and $\phi \in [0,\pi]$

Now we look at the time evolution of the components of the resulting Bloch vector of Eq.~\eqref{rbloch}. Note that $\mathbf{u}$ and $\mathbf{v}$ are orthogonal vectors, in such a way that $\{{\bf \hat{u}}, {\bf \hat{v}}, {\bf \hat{w}} \}$ forms an orthonormal basis for $\mathbb{R}^3$, where we have
\begin{equation}\label{w}
\mathbf{w}\equiv\mathbf{u}\times\mathbf{v}=\sin\theta\sin\phi\left({\bf \hat{x}}\cos\phi-{\bf \hat{z}}\cos\theta\sin\phi\right)
\end{equation}
 and ${\bf \hat{w}}\defeq\mathbf{w}/|\mathbf{w}|$. Substituting the above expressions for $\mathbf{u}$, $\mathbf{v}$ and $\mathbf{w}$ into Eqs.~\eqref{h}-\eqref{M}, the dynamics of the ${\bf \hat{u}}$, ${\bf \hat{v}}$, and ${\bf \hat{w}}$ components of $\mathbf{r}(t)$ is given by
\begin{eqnarray}\label{eq:dyn_comp_LI}
\mathbf{r}(t)\cdot{\bf \hat{u}}&=&\mathrm{e}^{-2\gamma_- t }(\mathbf{r}_0\cdot{\bf \hat{u}})\\
\mathbf{r}(t)\cdot{\bf \hat{v}}&=&\mathrm{e}^{-2\gamma_+ t }(\mathbf{r}_0\cdot{\bf \hat{v}})\\
\mathbf{r}(t)\cdot{\bf \hat{w}}&=&\mathrm{e}^{-4\gamma t}(\mathbf{r}_0\cdot{\bf \hat{w}})+2(1-\mathrm{e}^{-4\gamma t})|\mathbf{w}|\label{eq:dyn_comp_LIc}
\end{eqnarray}
where we have defined
\begin{equation}
2\gamma_\pm\defeq 2\gamma\pm\left[\cos(2\theta)\sin^2\phi+\cos^2\phi\right]^{1/2}\gamma\,.
\end{equation}

By comparing Eqs.~\eqref{eq:dyn_comp_LI} and~\eqref{eq:dyn_comp_LD}, two fundamental differences in the resulting dynamics of the Bloch vector are noticeable: First, in the present case, there is no longer an invariant component as the one implied by Eq.~\eqref{eq:dyn_comp_LDa}. Second, as time goes by, the second summand on the rhs of Eq.~\eqref{eq:dyn_comp_LIc} shifts the origin of the Bloch sphere along the ${\bf \hat{w}}$-direction by an amount proportional to $|\mathbf{w}|$, hence giving rise to a non-unital channel.

\begin{figure} [htbp]
\includegraphics[width=8.5cm]{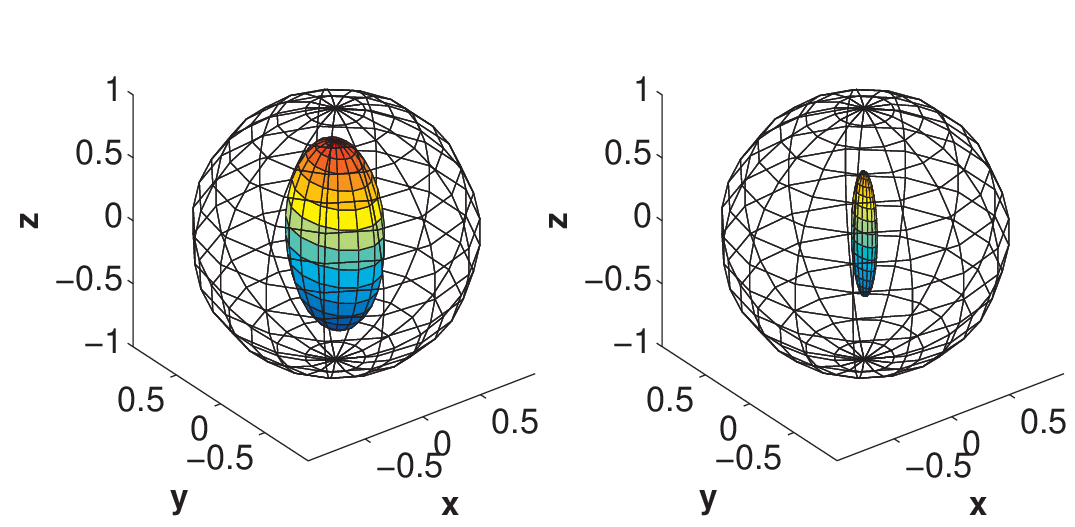}
\caption{\label{fig:depol_ch} Depolarizing channel. $\theta=\phi=0$. In the first Bloch sphere, $\gamma t = 0.3$, while in the second, $\gamma t = 0.7$.}
\end{figure}

In this regime, it is thus clear that unital channels can only be obtained if $|\mathbf{w}|=0$, or equivalently, if $\theta$ and $\phi$ satisfy either (i) $\sin\theta=0$ or (ii) $\sin\phi=0$ or (iii) $\cos\phi=0$ \emph{and} $\sin\theta=0$. However, in contrast to the previous section, the unital channels arising here require the lengths of all the Bloch vector components to exponentially decay with time, asymptotically collapsing the Bloch ball into a point corresponding to the maximally mixed state $\mathbb{I}_2/2$. A unital channel of this form is known as a depolarizing channel, and is illustrated in Fig.~\ref{fig:depol_ch} for two time steps.

\begin{figure} [htbp]
\includegraphics[width=8.5cm]{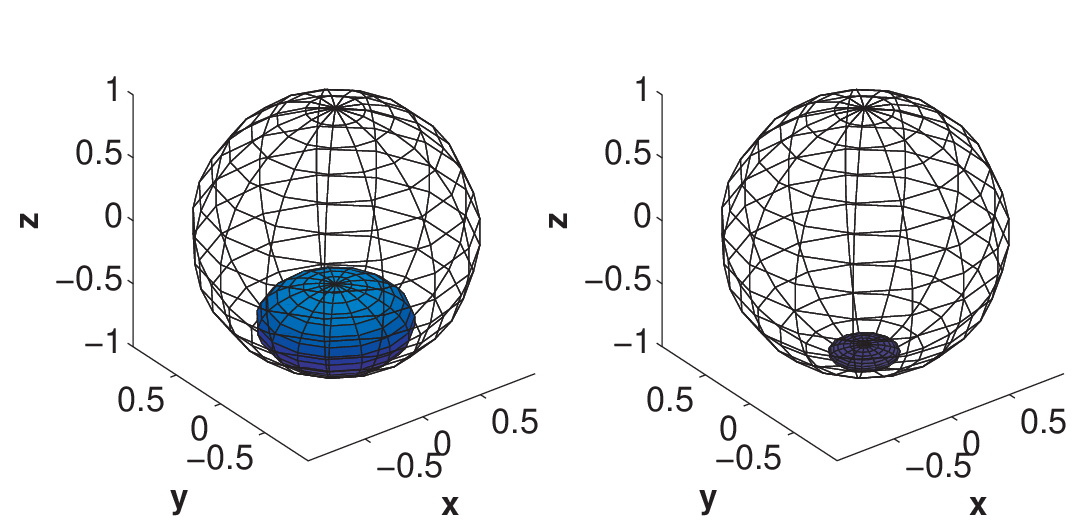}
\caption{\label{fig:ampdamp_ch} Amplitude-damping channel. $\theta=\pi/4$ and $\phi=\pi/2$. In the first Bloch sphere, $\gamma t = 0.3$, while in the second, $\gamma t = 0.7$.}
\end{figure}

As noted before, for $|\mathbf{w}|\neq 0$ a family of non-unital channels arises. A well-known representative of this family is the standard-amplitude-damping channel, occurring whenever $|\theta|=\pi/4$ and $\phi=\pi/2$. The action of this channel (with $\theta=+\pi/4$) on the Bloch sphere is illustrated in Fig.~\ref{fig:ampdamp_ch} for two time steps.

In analogy to the case of depolarizing channels, Fig.~\ref{fig:ampdamp_ch} illustrates that the Bloch ball asymptotically collapses into a single point. The difference between the two cases, though, is that for amplitude-damping this point corresponds to a pure state, as opposed to the maximally-mixed state in the case of the depolarizing channel. Clearly, the property of mapping the Bloch ball into a single point is nothing but a direct consequence of the limit establishing that $\mathbf{r}_\infty$ is independent of $\mathbf{r}_0$ for this type of interaction Hamiltonian (cf. Sec.~\ref{sec:asymptotics}).

Finally, it should be noted that generalized amplitude-damping channels can also be obtained by varying the parameters $\theta$ and $\phi$. In those more general cases, the Bloch sphere can collapse to states that are neither pure nor maximally mixed, as revealed by a quick inspection of the magnitude of the vector $\mathbf{w}$ of Eq.~\eqref{w}.

\section{Kraus Operators}\label{app:Kraus}
In this appendix, we show how to construct any set of Kraus operators for a channel acting on qudit states. Besides being of independent interest,  this result is useful to derive a set of Kraus operators to represent the dynamics of a single qubit evolving under a dissipative Hamiltonian (cf. Sec.~\ref{sec:dissipham}).

Before proving the main result, let us recall a few useful facts from the theory of quantum channels~\cite{04Zyczkowski3,05Salgado55,02Verstraete}.
Let $\mathcal{C}_d^{set}$ denote the set of all quantum channels $\mathcal{C}$ acting on $d$-dimensional density matrices, and let $\mathfrak{C}_{d^2}^{set}$ 
denote the set of all $d^2$-dimensional positive semidefinite matrices 
$\mathfrak{C}$
such that 
${\rm Tr}_2\mathfrak{C}=\mathbb{I}_d$.
It was independently shown in Refs.~\cite{99Fujiwara3290,99Horodecki1888} that $\mathcal{C}_d^{set}$ is isomorphic to 
$\mathfrak{C}_{d^2}^{set}$,
 with an isomorphism law --- sometimes referred to as the Choi-Jamio{\l}kowski isomorphism --- given by
\begin{equation}\label{eq:choi}
\mathfrak{C}=\sum_{j,k=1}^d\ket{j}\!\bra{k}\otimes\mathcal{C}(\ket{j}\!\bra{k})
\end{equation}
where $\{\ket{j}\}_{j=1}^d$ is an orthonormal basis for the Hilbert space of dimension $d$. Conversely, as it can be easily verified~\cite{01DAriano042308},
\begin{equation}
\mathcal{C(\rho)}={\rm Tr}_1\left[\left(\rho^{\sf T}\otimes \mathbb{I}_d\right)\mathfrak{C}\right]\,,
\end{equation}
where the transposition is taken with respect to the basis $\{\ket{j}\}_{j=1}^d$.

The crucial point to bear in mind is that the knowledge of a quantum channel $\mathcal{C}$ is entirely equivalent to the knowledge of the corresponding matrix 
$\mathfrak{C}$,
henceforth denominated the \emph{Choi matrix} of $\mathcal{C}$~\cite{75Choi285}. In what follows, we characterize the Kraus operators of $\mathcal{C}$ in terms of a decomposition of 
$\mathfrak{C}$.

\noindent
{\bf Lemma~2 [Verstraete~\cite{02Verstraete}, Salgado~\cite{05Salgado55}].}
{\it Let $\rho(t)=\mathcal{C}_t(\rho_0)$ be the density matrix of a qudit at time $t$, where $\mathcal{C}_t$ is a quantum channel and $\rho_0=\rho(0)$. Moreover, let 
$\mathfrak{C}(t)$
be the Choi matrix of $\mathcal{C}_t$, and consider any $d^2$-dimensional matrix $S(t)$ such that 
$S(t) S(t)^\dagger=\mathfrak{C}(t)$.
A Kraus decomposition for $\rho(t)$ is
 \begin{equation}
\rho(t)=\sum_i \left[{\rm mat}{S_i(t)}\right] \rho_0 \left[{\rm mat}{S_i(t)}\right]^\dagger
 \end{equation}
where $S_i(t)$ is the $i$-th column of $S(t)$, and ${\rm mat}{S_i(t)}$ is the $d\times d$ matrix whose $n$-th column is formed by the elements of $S_i$ with indexes ranging from $(n-1)d+1$ to $n d$.}

\noindent
{\bf Remark~1.}
The operation $mat$ is simply the inverse of the operation ${\rm vec}$~\cite{91Horn}, which vectorizes a matrix by stacking its columns. For example:
\begin{equation}\label{eq:vecmat_example}
{\rm vec}\left(\begin{array}{cc}a & c \\
b & d \end{array}\right)=\left(\begin{array}{c}a\\b\\c\\d\end{array}\right)\quad\mbox{and}\quad
{\rm mat}\left(\begin{array}{c}a\\b\\c\\d\end{array}\right)=\left(\begin{array}{cc}a & c \\
b & d \end{array}\right)\,.
\end{equation}

\noindent
{\bf Remark~2.} The Choi matrix of the quantum operations described in Sec.~\ref{sec:dissipham} can be easily computed via Eq.~\eqref{eq:choi} by noticing that $\mathcal{C}(\ket{0}\!\bra{0})$ and $\mathcal{C}(\ket{1}\!\bra{1})$ can be directly obtained from Eqs.~\eqref{rbloch}-\eqref{M} by making $\mathbf{r}_0=\mathbf{\hat{z}}$ and $\mathbf{r}_0=-\mathbf{\hat{z}}$, respectively. The matrices $\mathcal{C}(\ket{0}\!\bra{1})$ and $\mathcal{C}(\ket{1}\!\bra{0})$, in turn, can be obtained by exploiting the linearity of $\mathcal{C}$, explicitly
\begin{eqnarray}
2\mathcal{C}(\ket{0}\!\bra{1})&=&\mathcal{C}(\ket{+}\!\bra{+})-\mathcal{C}(\ket{-}\!\bra{-})\nonumber\\&+&i\mathcal{C}(\ket{+i}\!\bra{+i})-i\mathcal{C}(\ket{-i}\!\bra{-i})
\end{eqnarray}
\begin{eqnarray}
2\mathcal{C}(\ket{1}\!\bra{0})&=&\mathcal{C}(\ket{+}\!\bra{+})-\mathcal{C}(\ket{-}\!\bra{-})\nonumber\\&-&i\mathcal{C}(\ket{+i}\!\bra{+i})+i\mathcal{C}(\ket{-i}\!\bra{-i})
\end{eqnarray}
where $\mathcal{C}(\ket{\pm}\!\bra{\pm})$ and $\mathcal{C}(\ket{\pm i}\!\bra{\pm i})$ follow from Eqs.~\eqref{rbloch}-\eqref{M} by making $\mathbf{r}_0=\pm\mathbf{\hat{x}}$ and $\mathbf{r}_0=\pm\mathbf{\hat{y}}$, respectively.

\noindent
{\bf Remark~3.}
Since 
$\mathfrak{C}(t)$
 is a positive semidefinite matrix, it is always possible to find $S(t)$ such that 
 $S(t)S(t)^\dagger=\mathfrak{C}(t)$
 (e.g. Cholesky decomposition). In fact, there are (infinitely) many such choices of $S(t)$, which implies in (infinitely) many Kraus decompositions of $\rho(t)$~\cite{06Kirkpatrick95}.

\noindent
{\bf Proof of Lemma~2.}
The proof is a straightforward application of the following easy-to-check matrix identity:
\begin{equation}\label{eq:ABCvecid}
ABC^\dagger={\rm Tr}_1\left[(B^{\sf T}\otimes \mathbb{I}_d)({\rm vec}{A})({\rm vec}{C})^\dagger\right]
\end{equation}
for $A, B, C$ arbitrary $d$-dimensional matrices. In particular, take $A = C = {\rm mat}{S_i}$ and $B=\rho_0$, then
\begin{equation}\label{eq:vecidapplied}
\left[{\rm mat}{S_i}\right] \rho_0 \left[{\rm mat}{S_i}\right]^\dagger={\rm Tr}_1\left[\left(\rho_0^{\sf T}\otimes \mathbb{I}_d\right) S_i S_i^\dagger\right]
\end{equation}
where we have used that ${\rm vec}\left[{\rm mat} S_i\right] = S_i$. Summing over $i$ on both sides of Eq.~\eqref{eq:vecidapplied} and exploiting the linearity of the partial trace, the desired result follows:
\begin{eqnarray}
\sum_i \left[{\rm mat}{S_i}\right] \rho_0 \left[{\rm mat}{S_i}\right]^\dagger&=&{\rm Tr}_1\left[\left(\rho_0^{\sf T}\otimes \mathbb{I}_d\right) \sum_i S_i S_i^\dagger\right]\nonumber\\
&=&{\rm Tr}_1\left[\left(\rho_0^{\sf T}\otimes \mathbb{I}_d\right) S S^\dagger\right]\nonumber\\
&=&{\rm Tr}_1\left[\left(\rho_0^{\sf T}\otimes \mathbb{I}_d\right)\mathfrak{C}\right]\nonumber\\
&=&\mathcal{C}_t(\rho_0)\nonumber\\
&=&\rho(t).
\end{eqnarray}

\section{Watching SDE}\label{app:watch}
As an illustration of the result of Theorem~2, in this appendix we aim to visualize the occurrence of the SDE
for some instances of two-qubit entangled states, whose parties evolve independently under identical dissipative Hamiltonians.

For the initial conditions, we consider the states
\begin{eqnarray}
\ket{-}&\equiv&\alpha_{-}\ket{\uparrow\downarrow}+\beta_{-}\ket{\downarrow\uparrow},\label{anti}\\
\ket{+}&\equiv&\alpha_{+}\ket{\uparrow\uparrow}+\beta_{+}\ket{\downarrow\downarrow},\label{para}
\end{eqnarray}
where, for simplicity, we suppose $\alpha_\pm,\beta_\pm \in {\mathbb{R}}$.

For the Hamiltonians, instead of the usual absorption coupling agents $\sigma_{\pm}=(\sigma_x\pm i\sigma_y)/\sqrt{2}$ ---  or, in our notation, $\mathbf{u}^{(n)}=\mathbf{\hat{x}}/\sqrt{2}$ and $\mathbf{v}^{(n)}=\mathbf{\hat{y}}/\sqrt{2}$ --- we consider the more general dissipative couplings given by
\begin{equation}\label{eq:uv}
\mathbf{u}^{(n)}=\mathbf{\hat{x}}\cos{\theta}\qquad\mbox{and}\qquad \mathbf{v}^{(n)}=\mathbf{\hat{y}}\sin{\theta}\,,
\end{equation}
for $\theta \in [0,2\pi]$ and $n=1,2$.

\begin{figure}
\includegraphics[width=8.5cm]{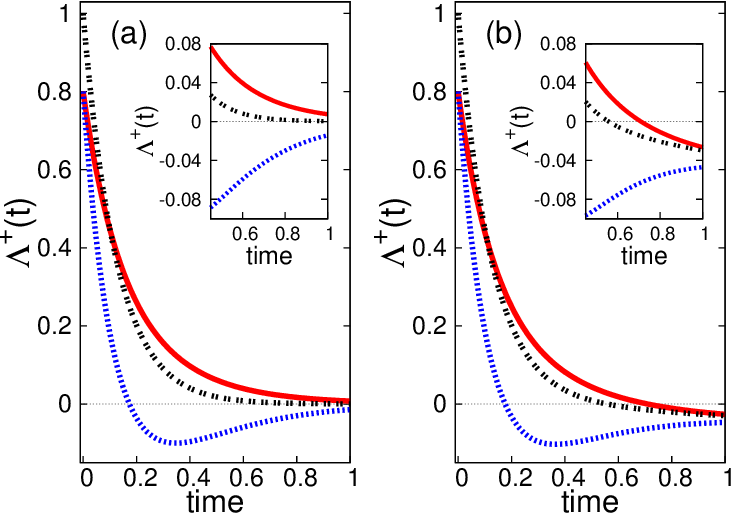}
\caption{\label{PARA} $\Lambda^+(t)$ for the initial conditions given by $|\alpha_+|^2=0.2$ for the red (solid) line, $|\alpha_+|^2=0.5$ for the black (double-dotted) line, and $|\alpha_+|^2=0.8$ for the blue (dotted) line. In (a) and (b) we use the interaction Hamiltonian as defined by Eq.~\eqref{eq:uv}, with $\theta=-\pi/4$ and $\theta=-\pi/5$, respectively.}
\end{figure}

\begin{figure} [htbp]
\includegraphics[width=8.5cm]{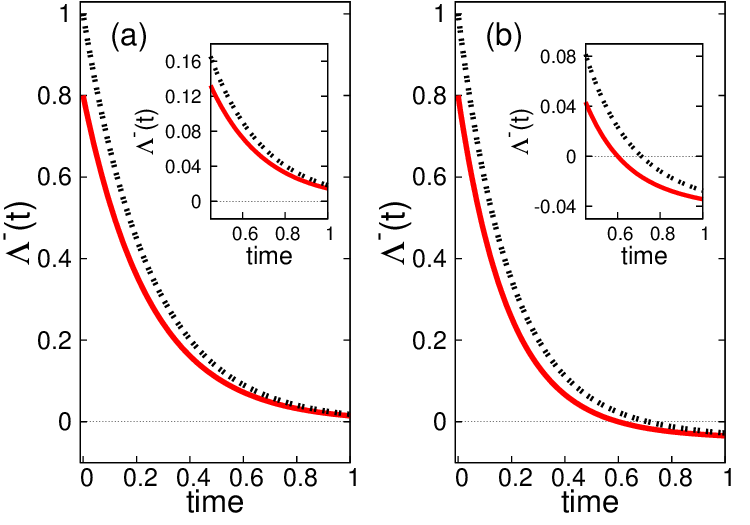}
\caption{\label{ANTI} $\Lambda^-(t)$ for the initial conditions given by $|\alpha_-|^2=0.2$ for the red (solid) line and $|\alpha_-|^2=0.5$ for the black (double-dotted) line (for this class of initial conditions, $|\alpha_-|^2=0.2$ and $|\alpha_-|^2=0.8$ give identical dynamics). In (a) and (b) we use the interaction Hamiltonian as defined by Eq.~\eqref{eq:uv}, with $\theta=-\pi/4$ and $\theta=-\pi/5$, respectively.}
\end{figure}

With the initial states and Hamiltonians thus specified, the dynamical equation~\eqref{eq:bellomo}, and the Kraus operators, it is possible to obtain the density-matrix dynamics for continuous time $t$. Furthermore, with a simple numerical procedure, the function $\Lambda(t)$ can also be computed in each case, and is plotted in Figs.~\ref{PARA} and~\ref{ANTI}.



Assuming $\gamma=1$ in Eq.~\eqref{lindblad} and the initial condition given by Eq.~\eqref{para},  we illustrate, in Fig. \ref{PARA}, $\Lambda^+(t)$ for $|\alpha_+|^2=0.2$, $|\alpha_+|^2=0.5$, and $|\alpha_+|^2=0.8$.
In Fig. \ref{PARA}$(a)$ we consider the usual amplitude-damping, supposing  $\theta=-\pi/4$ in Eq.~\eqref{eq:uv}, and we observe that the entanglement suddenly vanishes for $|\alpha_+|^2=0.8$. In fact, as it is already known in these circumstances~\cite{07Almeida579,08Aolita080501,08Salles022322,06Santos040305(R),07Yu459,06Yu140403,04Yu140404}, SDE occurs whenever $|\alpha_+|^2>0.5$. In Fig.~\ref{PARA}$(b)$ we consider $\theta=-\pi/5$ in Eq.~\eqref{eq:uv}. In this case there is no SDF subspace for the initial conditions of Eq.~\eqref{para}.

In Fig. \ref{ANTI} we consider the initial conditions given by Eq.~\eqref{anti} and keep all other variables as in Fig. \ref{PARA}. In this case, as illustrated in Fig. \ref{ANTI}$(a)$,  we have a SDF subspace for the usual amplitude-damping ($\theta=-\pi/4$). Otherwise, as Fig. \ref{ANTI}$(b)$ shows, SDE is present for all initial conditions when $\theta=-\pi/5$.

Finally, we note that the observations above are fully consistent with the statement of Theorem~2. The fact that SDE can occur even in the case of standard amplitude-damping channel proves that the condition of the theorem is not a necessary one. Moreover, the occurrence of SDE whenever a channel slightly different from the standard amplitude-damping is applied, supports the sufficiency clause.

\end{document}